\documentclass[aps,pra,twocolumn,superscriptaddress,showpacs]{revtex4}

\usepackage{float}
\usepackage{graphicx}
\usepackage{dcolumn}
\usepackage{amsmath}
\input{epsf}
\usepackage{epsfig}

\begin{document}

\title{Performance of various quantum key distribution systems using 1.55 $\mu$m up-conversion single-photon detectors}

\author{Eleni Diamanti}
\email{ediam@stanford.edu} \affiliation{Edward L. Ginzton
Laboratory, Stanford University, Stanford, California 94305, USA}
\author{Hiroki Takesue}
\affiliation{NTT Basic Research Laboratories, NTT Corporation,
Kanagawa 243-0198, Japan}
\author{Toshimori Honjo}
\affiliation{NTT Basic Research Laboratories, NTT Corporation,
Kanagawa 243-0198, Japan}
\author{Kyo Inoue}
\affiliation{NTT Basic Research Laboratories, NTT Corporation,
Kanagawa 243-0198, Japan}
\author{Yoshihisa Yamamoto}
\affiliation{Edward L. Ginzton Laboratory, Stanford University,
Stanford, California 94305, USA} \affiliation{NTT Basic Research
Laboratories, NTT Corporation, Kanagawa 243-0198, Japan}

\date{\today}

\begin{abstract}
We compare the performance of various quantum key distribution (QKD)
systems using a novel single-photon detector, which combines
frequency up-conversion in a periodically poled lithium niobate
(PPLN) waveguide and a silicon avalanche photodiode (APD). The
comparison is based on the secure communication rate as a function
of distance for three QKD protocols: the Bennett-Brassard 1984
(BB84), the Bennett, Brassard, and Mermin 1992 (BBM92), and the
coherent differential phase shift keying (DPSK). We show that the
up-conversion detector allows for higher communication rates and
longer communication distances than the commonly used InGaAs/InP APD
for all the three QKD protocols.
\end{abstract}


\maketitle

\section{\label{sec:intro}Introduction}
Quantum key distribution (QKD) allows two parties to share an
unconditionally secure secret key. Security is guaranteed by the
laws of quantum mechanics, ensuring that the key can be used
afterwards to encrypt and decrypt secret messages as a one-time
pad. The most common QKD protocols, which have been implemented in
experiments over the last years \cite{gisin}, are the BB84
protocol, which uses single photons as information carriers
\cite{bb84}, and the entanglement-based BBM92 protocol
\cite{bbm92}. A security analysis for these protocols under
realistic system parameters and against individual attacks has
been performed \cite{lutkenhaus,waks}. This analysis shows that
the performance of a quantum cryptography system, in terms of
communication distance and secure communication rate, is
determined by the characteristics of the source of single or
entangled photons, and of the single-photon detectors. In addition
to the BB84 and BBM92 protocols, we consider the recently proposed
differential phase shift keying (DPSK) protocol, which uses a weak
coherent pulse train as the information carrier
\cite{inoue-dpsk,inoue-dpskc}. To this end, we develop a security
analysis against certain types of hybrid attacks.

To date, fiber-optic QKD systems have invariably used InGaAs/InP
avalanche photodiodes (APDs) as single-photon detectors. Recently,
an alternative technology for very efficient single-photon detection
at 1.55 $\mu$m, based on the principle of frequency up-conversion,
was presented \cite{langrock}. Using realistic experimental
parameters, we perform comparisons for the various types of sources
and protocols, and show that longer communication distances and
higher communication rates can be achieved using the up-conversion
detector in all cases.

\section{\label{sec:det}1.55 $\bf{\mu m}$ single-photon detectors}

\subsection{\label{sec:APD}InGaAs/InP avalanche photodiode}
The InGaAs/InP avalanche photodiodes have been the subject of
thorough investigation over the last decade due to their importance
as single-photon detectors in fiber-optic QKD implementations.
Although considerable progress has been achieved in the performance
of these detectors \cite{yoshizawa,bethune,stucki,bourennane,gobby},
they exhibit low quantum efficiencies (typically on the order of
0.1), and, most seriously, they suffer from after-pulse effects
caused by trapped charge carriers, which produce large dark count
rates during a relatively long time. The high dark count probability
imposes \emph{gated-mode operation}, which limits their capabilities
significantly. In particular, when operated in gated mode, the APD
device is raised above breakdown threshold for a few nsec, which
ensures low probability of a dark count and high efficiency for
detecting light. Subsequently, the device is returned to below
breakdown for a time long enough for any trapped charge carrier to
leak away. Given that the trapping lifetime is on the order of a
$\mu$sec, this mode allows operation at MHz rates, while the
after-pulse probability is reduced by the ratio of the gate width to
the time separation between gates. In a QKD application, this gate
frequency determines the repetition rate of the signal pulse and,
therefore, limits the attainable communication rate. Furthermore,
the dark count rate, which is critical for the communication
distance, is determined by the gate width, limited by the response
time of the semiconductor material. Typically, gate widths of $1-2$
nsec at $\sim 1$ MHz repetition frequency are used with resulting
dark count rates on the order of $10^4$/sec.

\begin{figure}[t]
\epsfxsize=3.25in \epsfbox{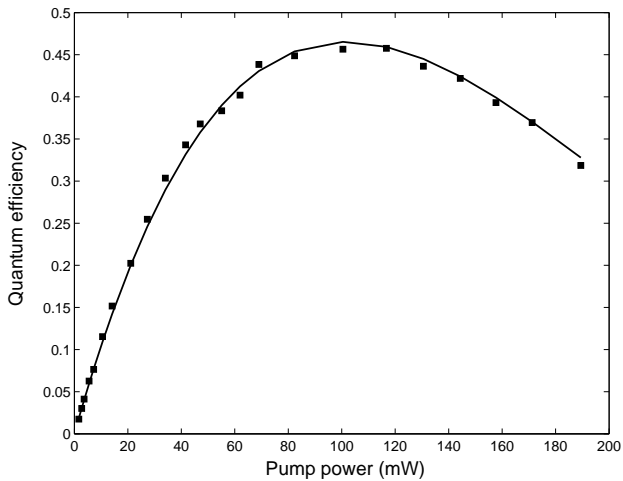} \caption{Quantum efficiency of
the 1.55 $\mu$m up-conversion single photon detector as a function
of pump power. The expression for the fitting curve is given by
Eq.~(\ref{eq:etaup}).} \label{fig:detectoreta}
\end{figure}

\subsection{\label{sec:up}Up-conversion detector}
In the 1.55 $\mu$m up-conversion single-photon detector
\cite{langrock}, a single photon at 1.55 $\mu$m interacts with a
strong pump at 1.32 $\mu$m in a periodically poled lithium niobate
(PPLN) waveguide, designed for sum-frequency generation at these
wavelengths \cite{roussev}. Due to the quasi-phase-matching and the
tight mode confinement over long interaction lengths achieved in a
guided-wave structure, this device allows for very high conversion
efficiency of the signal to the $\sim$ 0.7 $\mu$m sum frequency
output. The converted photon is subsequently detected by a silicon
APD. Contrary to InGaAs/InP APDs, Si APDs have high quantum
efficiencies in the near-infrared (typically on the order of
$0.6-0.7$), very low dark count rates, and very small after-pulse
effects. The last characteristic enables \emph{Geiger (non-gated)
mode operation} of the Si APD, which does not impose any severe
limitation to the attainable communication rate in a QKD system. In
practice, however, the rate is limited by the dead time of Si APD
detectors, which is on the order of 50 nsec for commercial devices.
During this time period that follows a photo-detection event, the
photodiode cannot respond to subsequent events, and, eventually, a
very large photon flux saturates the device. This effect is taken
into account in the calculations of Section \ref{sec:calc}.

\begin{figure}[t]
\epsfxsize=3in \epsfbox{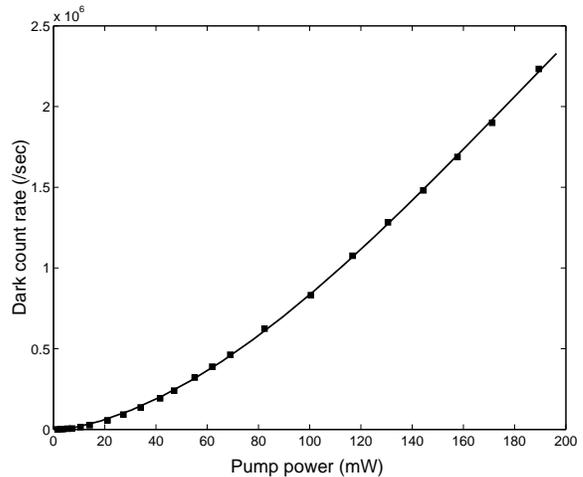} \caption{Dark count rate of the
1.55 $\mu$m up-conversion single photon detector as a function of
pump power. The expression for the fitting curve is given by
Eq.~(\ref{eq:dcup}).} \label{fig:detectordc}
\end{figure}

The main characteristics of the up-conversion detector, such as the
quantum efficiency, $\eta_{\text{up}}$, and the dark count rate,
$D_{\text{up}}$, depend on the pump power, $p$ \cite{langrock}. When
the phase-matching condition in the waveguide is met and sufficient
pump power is available to achieve almost 100\% photon conversion, a
maximum overall quantum efficiency of 0.46 is achieved, as shown in
Fig.~\ref{fig:detectoreta}. In agreement with the coupled mode
theory for three-wave interactions in a waveguide, which predicts a
$\text{sin}^2$ dependence of $\eta_{\text{up}}$ on $p$, the fitting
curve of the experimental results is given by the following
expression:
\begin{equation}
\eta_{\text{up}}(p) = a_1 \sin^2{(\sqrt{a_2 p})} \label{eq:etaup}
\end{equation}
where $a_1 = 0.465, a_2 = 79.75$, and $p$ is given in mW.

On the other hand, the dark count rate is dominated by a combined
nonlinear process: Initially, the pump photons are scattered by the
phonons of both the PPLN waveguide and the fiber via a spontaneous
Raman scattering process. This process scales linearly with the pump
power, and generates a spectrum of Stokes photons, which includes
the signal wavelength of 1.55 $\mu$m. Subsequently, the noise
photons interact with the pump photons in the waveguide via the
phase-matched sum-frequency generation process, and create dark
counts. The combined process results in an approximately quadratic
dependence of the dark counts on the pump power, as shown in
Fig.~\ref{fig:detectordc}. A more accurate polynomial fitting curve
is given by the following expression:
\begin{equation}
D_{\text{up}}(p) = b_0 + b_1 p + b_2 p^2 + b_3 p^3 + b_4 p^4 \mbox{
}\mbox{ }\mbox{ } \text{(/sec)} \label{eq:dcup}
\end{equation}
where $b_0=50$, $b_1=826.4$, $b_2=110.3$, $b_3=-0.403$,
$b_4=0.00065$, and $p$ is again given in mW.

\begin{figure}[b]
\epsfxsize=3.25in \epsfbox{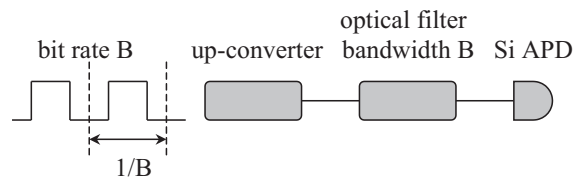} \caption{Ideal communication
system employing an up-conversion detector.}
\label{fig:ideal-system}
\end{figure}

\begin{figure}[t]
\epsfxsize=3.25in \epsfbox{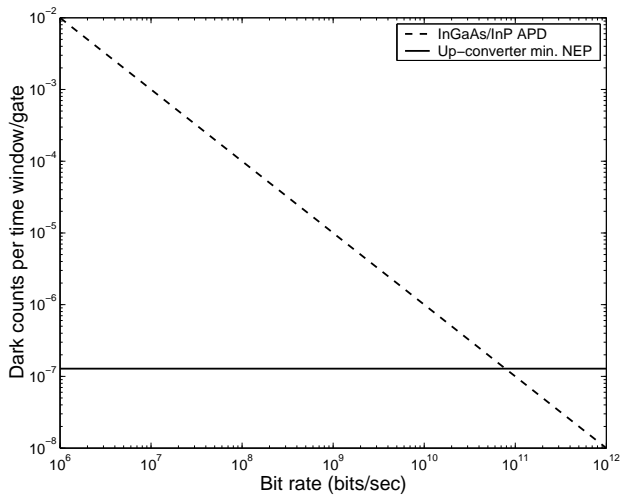} \caption{Dark counts per time
window/gate for the up-conversion single-photon detector operating
at the minimum NEP regime, and a typical InGaAs/InP APD
respectively, in the communication system described in
Fig.~\ref{fig:ideal-system}.} \label{fig:dgate}
\end{figure}

An important feature of the up-conversion detector stems from the
fact that the dark counts depend on the bandwidth of the waveguide,
as this determines the number of noise photons. We can define a
quantity, $D_{\text{up-Hz}}=\frac{D_{\text{up}}}{B_{\text{d}}}$
$(\text{sec}^{-1}\text{Hz}^{-1})$, for a detector with bandwidth
$B_{\text{d}}$, which corresponds to the dark counts per mode. Then,
we can think of the ideal communication system shown in
Fig.~\ref{fig:ideal-system} with a matched filter with bandwidth
equal to the bit rate $B$. In such a system, the dark counts per
time window, $d_{\text{up}}$, a parameter of great importance in QKD
applications, is equal to $D_{\text{up-Hz}}$. Note that
$d_{\text{up}}$ is independent of the bit rate $B$ (or measurement
time window $1/B$) under this optimum filtering. An InGaAs/InP APD
operated in gated mode has dark counts per gate, $d_{\text{APD}}$,
calculated by $D_{\text{APD}}\frac{1}{B}$, where $D_{\text{APD}}$
(/sec) is the dark count rate of the InGaAs/InP APD. In
Fig.~\ref{fig:dgate}, the quantity $d$ is plotted for the two types
of detectors as a function of the bit rate. For the InGaAs/InP APD,
the typical value $D_{\text{APD}}=10^4$/sec is used. For the
up-conversion detector, we calculate the quantity $D_{\text{up-Hz}}$
at the operating point of the detector, where the normalized Noise
Equivalent Power (NEP), $\sqrt{2 D}/\eta$, is minimized, which
corresponds to $D_{\text{up}}=6.4\times 10^3$/sec and
$\eta_{\text{up}}=0.075$. Given a bandwidth of $B_{\text{d}}=50$ GHz
for the up-converter, we find that the optimum $d_{\text{up}}$ is
$\sim 1.3\times 10^{-7}$, as shown in Fig.~\ref{fig:dgate}. This
result illustrates the significant advantage of the up-conversion
detector for most practical system bit rates.

The dependence of the dark counts on the waveguide bandwidth,
together with the non-gated mode operation of the Si APD and the
pump power dependence of the detector characteristics, have a
significant effect on the performance of a quantum cryptography
system employing up-conversion detectors, as we will see in the
following sections.

\section{\label{sec:rate}Communication rate equations}
In this paper, we will consider only individual attacks, that is Eve
is restricted to attack only individual bits; she is not allowed to
perform a coherent attack consisting of collective quantum
operations and measurements of many qubits with quantum computers.
In a QKD system, the raw transmission of random bits is followed by
a public exchange of information on the time of single-photon
detection and the bases used by the two parties, which results in
the \emph{sifted key}. The steps of classical error-correction and
privacy amplification follow. The first step serves the dual purpose
of correcting all erroneously received bits and giving an estimate
of the error rate. Privacy amplification is then used to distill a
shorter key, the \emph{final key}, which can be made as secure as
desired. The security analyses of \cite{lutkenhaus,waks} take all
the above steps into account and derive the communication rate
equations that are re-stated in Sections~\ref{sec:BB84} and
\ref{sec:BBM92}. In Section~\ref{sec:DPS}, we derive the
corresponding equation for the DPSK protocol, based on the security
analysis against certain types of hybrid attacks.

\subsection{\label{sec:BB84}BB84 protocol}
In the BB84 protocol, Alice sends Bob single photons randomly
modulated in two non-orthogonal bases. Bob measures the polarization
states of the single photons in a randomly chosen polarization
basis. The secure communication rate of this protocol against an
arbitrary individual attack, including the most commonly considered
intercept-resend and photon-number splitting (PNS) attacks
\cite{lutkenhaus}, is given by the following expression:
\begin{eqnarray}
R_{\text{BB84}}&=&\frac{1}{2}\nu p_{\text{click}}\{\tau(e,\beta)+f(e)[e\log_2 e+ \nonumber \\
&& (1-e)\log_2(1-e)]\} \label{eq:RBB84}
\end{eqnarray}
In the above equation, the factor $\frac{1}{2}$ is called the
sifting parameter and is due to the fact that half of the times
Alice's and Bob's polarization bases are not the same. The
repetition rate of the transmission is given by $\nu$. The
probability that Bob detects a photon is
\begin{equation}
p_{\text{click}}=p_{\text{signal}}+p_{\text{dark}}
\label{eq:pclickBB84}
\end{equation}
Simultaneous signal and dark counts are ignored in the above
expression, and the two components are given by
\begin{eqnarray}
p_{\text{signal}}&=&\mu \eta 10^{-(\alpha L+L_{\text{r}})/10} \label{eq:psignalBB84} \\
p_{\text{dark}}&=&4 d \label{eq:pdarkBB84}
\end{eqnarray}
where $\mu$ is the average number of photons per pulse, $\eta$ the
quantum efficiency of the detector, $\alpha$ the loss coefficient of
the optical fiber in dB/km, $L$ the distance in km, $L_{\text{r}}$
the loss of the receiver unit in dB, and $d$ the dark counts per
measurement time window of the system. The coefficient 4 in
Eq.~(\ref{eq:pdarkBB84}) is due to the assumption of a passive
detection unit involving four detectors at Bob's site, as in
\cite{waks}. For an ideal single-photon source, $\mu = 1$, while for
a Poisson source, which corresponds to the common weak laser pulse
implementations \cite{gisin}, $\mu$ becomes a free variable which
should be optimized.

The error rate is given by the expression:
\begin{equation}
e = \frac{\frac{1}{2}p_{\text{dark}}+ b
p_{\text{signal}}}{p_{\text{click}}} \label{eq:eBB84}
\end{equation}
where $b$ is the baseline system error rate, which cannot be
distinguished from tampering. The last term in Eq.~(\ref{eq:RBB84})
corresponds to the additional shrinking of the sifted key due to the
leakage of information to Eve during classical error correction. The
function $f(e)$ depends on the error-correction algorithm and its
values are given in Table~\ref{tab:fe} for the bi-directional
algorithm developed in \cite{salvail}.

\begin{table}[t]
\caption{\label{tab:fe}Benchmark performance of the error-correction
algorithm given in \cite{salvail}.}
\begin{ruledtabular}
\begin{tabular}{cp{1in}}
\hspace{0.7in}$e$ & $f(e)$ \\
\hline \\
\hspace{0.7in}0.01 & 1.16 \\
\hspace{0.7in}0.05 & 1.16 \\
\hspace{0.7in}0.1 & 1.22 \\
\hspace{0.7in}0.15 & 1.35
\end{tabular}
\end{ruledtabular}
\end{table}

Finally, the main shrinking factor $\tau(e,\beta)$ in the privacy
amplification step is related through the expression
\begin{equation}
\tau = -\log_2 p_{\text{c}} \label{eq:taupc}
\end{equation}
to the average collision probability, $p_{\text{c}}$. This is a
measure of Eve's mutual information with Alice and Bob. In
\cite{lutkenhaus} the following result is derived for $\tau$:
\begin{equation}
\tau(e,\beta) =
-\beta\log_2\left[\frac{1}{2}+2\frac{e}{\beta}-2\left(\frac{e}{\beta}\right)^2\right]
\label{eq:tauBB84}
\end{equation}
The parameter $\beta$ is defined as the fraction of single-photon
states emitted by the source:
\begin{equation}
\beta = \frac{p_{\text{click}}-p_{\text{m}}}{p_{\text{click}}}
\label{eq:betaBB84}
\end{equation}
where $p_{\text{m}}$ is the probability that the source emits a
multi-photon state. For an ideal single-photon source, $p_{\text{m}}
= 0$ (i.e., $\beta = 1$), while for a Poisson source,
\begin{equation}
p_{\text{m}} = 1-(1+\mu)e^{-\mu} \label{eq:pm}
\end{equation}
Essentially, the parameter $\beta$ accounts for the PNS attacks,
with which Eve can obtain full information without causing any error
in the communication between Alice and Bob by performing a quantum
non-demolition (QND) measurement of the photon number in each pulse,
keeping one photon in her quantum memory when she detects multiple
photons, and applying a delayed measurement on her photon after the
public announcement of the bases by Bob. This attack is a major
restricting factor in the performance of a weak laser pulse
implementation of the BB84 protocol. The secure communication rate
decreases quadratically with the transmission of the quantum
channel, $10^{-\alpha L/10}$, for small error rate and
$p_{\text{dark}}\ll p_{\text{signal}} \ll 1$. On the contrary, for
an ideal single-photon source implementation, under the same
conditions we find $R_{\text{BB84}}\approx \frac{1}{2}\nu
p_{\text{signal}}$, i.e., the rate decreases only linearly with the
fiber transmission.

The above security analysis is based on the assumption that Eve has
a quantum memory with an infinitely long coherence time because
Alice and Bob can delay the public announcement for an arbitrarily
long time. If Eve is not equipped with such a quantum memory, she
must perform the polarization measurement with a randomly chosen
basis. In this realistic case, Eq.~(\ref{eq:tauBB84}) must be
modified to:
\begin{equation}
\tau(e,\beta) =
-\frac{1+\beta}{2}\log_2\left[\frac{1}{2}+4\frac{e}{1+\beta}-8\left(\frac{e}{1+\beta}\right)^2\right]
\label{eq:tauBB84nomem}
\end{equation}

\subsection{\label{sec:BBM92}BBM92 protocol}
The BBM92 protocol is the two-photon variant of BB84. Alice and Bob
each share a photon of an entangled photon-pair, for which they
measure the polarization state in a randomly-chosen basis out of two
non-orthogonal bases. It was shown in \cite{waks} that the average
collision probability, $p_{\text{c}}$, for this protocol is the same
as that of the BB84 with a single-photon source, i.e., with
$\beta=1$. The shrinking factor $\tau$ becomes:
\begin{equation}
\tau(e) = -\log_2\left[\frac{1}{2}+2e-2e^2\right]
\label{eq:tauBBM92}
\end{equation}
This indicates that there is no analog  to a photon-number splitting
attack in BBM92. In general, the nature of this entanglement-based
protocol renders it more robust than BB84; for example it is less
vulnerable to errors caused by dark counts, since one dark count
alone cannot produce an error in this protocol. The equation for the
secure communication rate against any individual attack is given by
the following expression \cite{waks}:
\begin{eqnarray}
R_{\text{BBM92}}&=&\frac{1}{2}\nu p_{\text{coin}}\{\tau(e)+f(e)[e\log_2 e+ \nonumber \\
&& (1-e)\log_2(1-e)]\} \label{eq:RBBM92}
\end{eqnarray}
The sifting parameter is the same as in BB84, while the probability
of a coincidence between Alice and Bob is
\begin{equation}
p_{\text{coin}}=p_{\text{true}}+p_{\text{false}}
\label{eq:pcoinBBM92}
\end{equation}
The expressions for the probability of a true coincidence,
$p_{\text{true}}$, and the probability of a false coincidence,
$p_{\text{false}}$, are different for a deterministic
entangled-photon source and a Poissonian entangled-photon source,
such as a parametric down converter (PDC). They are given below,
under the assumption that the source is placed halfway between the
two parties \cite{waks}.
\begin{enumerate}
\item Deterministic entangled-photon source
\begin{eqnarray}
p_{\text{true}}&=&\eta^2 10^{-(\alpha L+2L_{\text{r}})/10} \label{eq:ptrueideal} \\
p_{\text{false}}&=&8 d \eta 10^{-(\alpha L+2L_{\text{r}})/20}+16 d^2
\label{eq:pfalseideal}
\end{eqnarray}
\item Poissonian entangled-photon source
\begin{eqnarray}
p_{\text{true}}&=&c_1 \label{eq:ptruePDC} \\
p_{\text{false}}&=&16 d^2 c_2 + 8 d c_3 + c_4 \label{eq:pfalsePDC}
\end{eqnarray}
where
\begin{eqnarray}
c_1&=&\frac{1}{\cosh^4\chi}\frac{2 t_L^2\tanh^2\chi}
{\left[1-\tanh^2\chi(1-t_L)^2\right]^4} \label{eq:c1} \\
c_2&=&\frac{1}{\cosh^4\chi}
\frac{1}{\left[1-\tanh^2\chi(1-t_L)^2\right]^2} \label{eq:c2} \\
c_3&=&\frac{1}{\cosh^4\chi}\frac{2 t_L(1-t_L)\tanh^2\chi}
{\left[1-\tanh^2\chi(1-t_L)^2\right]^3} \label{eq:c3} \\
c_4&=&\frac{1}{\cosh^4\chi}\frac{4 t_L^2(1-t_L)^2\tanh^4\chi}
{\left[1-\tanh^2\chi(1-t_L)^2\right]^4} \label{eq:c4}
\end{eqnarray}
and
\begin{equation}
t_L = \eta 10^{-(\alpha L+2L_{\text{r}})/20} \label{eq:tL}
\end{equation}
\end{enumerate}

All the parameters in the above equations are defined as in the
previous section. The parameter $\chi$, which appears in the case of
the Poissonian entangled-photon source, is a free variable that
depends on the average photon-pair number per pulse, i.e. the
nonlinear coefficient, the pump energy and the interaction time of
the down conversion process. Finally, the error rate is given by the
expression:
\begin{equation}
e = \frac{\frac{1}{2}p_{\text{false}}+ b
p_{\text{true}}}{p_{\text{coin}}} \label{eq:eBBM92}
\end{equation}

For small error rate and $p_{\text{false}}\ll p_{\text{true}}$, the
secure communication rate of BBM92 decreases linearly with the
transmission of the quantum channel, similarly to the case of the
BB84 protocol with a single photon source. Note that Eve does not
need a quantum memory to attack the BBM92 protocol. Equation
(\ref{eq:RBBM92}) is solely determined by the intercept and resend
attack.

\subsection{\label{sec:DPS}DPSK protocol}
Instead of using two non-orthogonal bases as in BB84 and BBM92,
the differential phase shift keying (DPSK) protocol uses many
non-orthogonal states consisting of many pulses
\cite{inoue-dpsk,inoue-dpskc}. In particular, it is based on the
fact that highly attenuated coherent states of many pulses with
random $\{0,\pi\}$ phase modulation are mutually non-orthogonal.
The idea of encoding the information in the phase of highly
attenuated coherent pulses was first presented by Bennett in 1992
\cite{b92}. The DPSK protocol is a simpler but more efficient
protocol compared to the B92 protocol. A similar protocol has also
recently been proposed \cite{gisin2}.

\begin{figure}[t]
\epsfxsize=3.5in \epsfbox{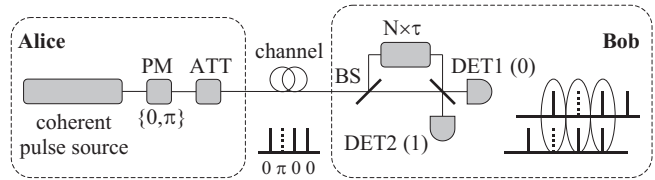} \caption{Configuration of the
DPSK protocol. PM: phase modulator, ATT: attenuator, BS:
beam-splitter, DET: detector.} \label{fig:dps-system}
\end{figure}

In the DPSK protocol, shown in Fig.~\ref{fig:dps-system}, all pulses
are highly attenuated and randomly phase-modulated by $\{0,\pi\}$.
Each photon coherently spreads over many pulses with a fixed phase
modulation pattern. In the receiver side, Bob randomly modulates the
delay time $N\times\tau$ in his interferometer by randomly choosing
a positive integer $N$, as shown in Fig.~\ref{fig:dps-system}, where
$\tau$ is the inverse of the clock frequency. After passing through
Bob's interferometer, the pulses interfere at Bob's output
beam-splitter, and which detector clicks depends on the phase
difference of the two pulses separated by a time $N\times\tau$. Bob
announces publicly the time instances at which a photon was detected
and the randomly chosen positive integer $N$. From her modulation
data Alice knows which detector recorded the event. Thus, they form
a secret key by assigning a bit value to each detector. The shifting
parameter is 1 since all bits are utilized during the key formation.

The security of the DPSK protocol stems from the fact that the
information is encoded on the differential phase of two nonlocal
pulses. This renders the protocol robust against any type of
individual photon splitting attack \cite{inoue-prasec,honjo}. In
order to derive the communication rate equation, we need to
calculate the privacy amplification shrinking factor, $\tau$,
defined in Eq.~(\ref{eq:taupc}) as a function of the average
collision probability, $p_{\text{c}}$. Our analysis takes into
account a
hybrid attack, which consists of two types of collective attacks:\\

\hspace{0.01in}1. Beam-splitter attack \\

Eve uses a beam-splitter with transmission $\eta_{\text{BS}}$ to
obtain coherent copies of the quantum state of many pulses that
Alice sends to Bob. She also replaces the lossy optical fiber with a
loss-less one, and the inefficient detectors at Bob's receiver unit
with ideal ones. Without Eve's intervention, Bob's probability of
detecting a signal photon, $p_{\text{signal}}$, is identical to the
one given in Eq.~(\ref{eq:psignalBB84}). In order to leave this
probability unaltered, Eve has to set the beam-splitter
transmission, $\eta_{\text{BS}}$, to:
\begin{equation}
\eta_{\text{BS}}=\eta 10^{-(\alpha L+L_{\text{r}})/10}
\label{eq:etaBS}
\end{equation}
where all the parameters are defined as in Section~\ref{sec:BB84}.
One possibility for Eve is to measure the pulses that she picks up
with an interferometer with delay time $M\times\tau$ chosen
independently from Bob's. In this case, her information gain is
calculated as follows: the probability of a detection event at Eve's
and Bob's site at a given time slot is given by
$\mu(1-\eta_{\text{BS}})$ and $\mu\eta_{\text{BS}}$ respectively,
where $\mu$ is the average number of photons per pulse. Thus, the
probability of a detection event at the same time instance is equal
to $\mu^2\eta_{\text{BS}}(1-\eta_{\text{BS}})$. On the other hand,
the probability that Eve's randomly chosen $M$ matches Bob's $N$ is
equal to $1/N$. Then, the probability that Eve gains bit information
relative to Bob is
$\mu^2\eta_{\text{BS}}(1-\eta_{\text{BS}})/(\mu\eta_{\text{BS}}N)=\mu(1-\eta_{\text{BS}})/N$.
This is true if we assume that Eve is not equipped with a quantum
memory with an infinitely long coherence time or if Alice and Bob
encrypt their public channel communication. However, if we allow Eve
to have a quantum memory and the two parties do not encrypt their
public exchange of information, Eve's strategy can be changed in
order to increase her information gain. In this case, she keeps the
pulses in her quantum memory and waits for Bob's announcement. Note
that Alice and Bob can delay the public announcement for an
arbitrarily long time, so Eve's quantum memory must have an
infinitely long coherence time. Then, Eve uses an optical
interferometer with an active switch that allows her to interfere
only the pulses for which she is aware that Bob has obtained the
differential phase information. This strategy increases Eve's
probability of gaining bit information to
$2\mu(1-\eta_{\text{BS}})$. The beam-splitter attack does not cause
any error in the communication between Alice and Bob, hence it gives
full information, i.e., $p_{\text{c}}=1$, to Eve for a fraction of
bits equal to $\mu(1-\eta_{\text{BS}})/N$ or
$2\mu(1-\eta_{\text{BS}})$. The remaining fraction of the bits is
given by:
\begin{equation}
\gamma = \left\{ \begin{array}{ll}
          1-\frac{\mu(1-\eta_{\text{BS}})}{N}= 1-\frac{\mu}{N}+\frac{p_{\text{signal}}}{N}\\
          \mbox{ }\mbox{ }\mbox{ }\mbox{ }\mbox{ }\mbox{ }\mbox{ }\mbox{ }\mbox{ }\mbox{ }
          \mbox{ }\mbox{ }\mbox{ }\mbox{ }\mbox{ }\text{:without quantum memory}\\
          1-2\mu(1-\eta_{\text{BS}})= 1-2\mu+2p_{\text{signal}}\\
          \mbox{ }\mbox{ }\mbox{ }\mbox{ }\mbox{ }\mbox{ }\mbox{ }\mbox{ }\mbox{ }\mbox{ }
          \mbox{ }\mbox{ }\mbox{ }\mbox{ }\mbox{ }\text{:with quantum memory}
          \end{array} \right.
\label{eq:gamma}
\end{equation}

\vspace{0.08in}

\hspace{0.01in}2. Intercept-resend attack \\

Eve also applies an intercept and resend attack to some of the
pulses that are sent to Bob after her beam splitter. In particular,
Eve intercepts two pulses with a time interval $M\times\tau$, lets
them pass through an interferometer with an identical delay
$M\times\tau$, measures the differential phase, and according to her
measurement result she sends an appropriate state to Bob. We assume
that in the case of an inconclusive or vacuum outcome she sends the
vacuum state, while when she measures a single photon she sends a
photon split into two pulses with the correct phase difference
applied between them. In this case, when Bob picks up an identical
delay, $N=M$, and measures the central time slot, he does not detect
the eavesdropping because he obtains the correct answer. However,
with probability $1-\frac{1}{2N}$ he chooses another delay, $N\neq
M$, or measures the side time slots, which yield random,
uncorrelated results, and with probability $\frac{1}{2}$ these lead
in error. Hence, this attack causes a bit error of
$\frac{1}{2}\left(1-\frac{1}{2N}\right)$ in the communication
between Alice and Bob. If the error rate of the system is $e$, Eve
is allowed to apply her attack to a fraction $\frac{2e}{1-1/2N}$ of
the pulse-pairs in order not to exceed this error rate. With
probability $\frac{1}{2N}$, she obtains full information for these
intercepted pulse-pairs.\\

In summary, taking into account the hybrid attack consisting of the
beam-splitter and intercept-resend attacks, we find that the
fraction of bits for which Eve has no information, i.e., for which
$p_{\text{c}}=\frac{1}{2}$, is equal to
$\gamma-\frac{e}{N(1-1/2N)}$. Thus, we have calculated the privacy
amplification shrinking factor,
\begin{equation}
\tau(e,\gamma)=\gamma-\frac{e}{N\left(1-\frac{1}{2N}\right)}
\label{eq:tauDPS}
\end{equation}
where $\gamma$ is given by Eq.~(\ref{eq:gamma}). We can now write
the equation for the secure communication rate of the DPSK protocol
against the hybrid attack we considered:
\begin{eqnarray}
R_{\text{DPSK}}&=&\nu p_{\text{click}}\{\tau(e,\gamma)+f(e)[e\log_2 e+ \nonumber \\
&& (1-e)\log_2(1-e)]\} \label{eq:RDPS}
\end{eqnarray}
In the above equation, $\nu$ is the repetition rate of the
transmission. The probability that Bob detects a photon,
$p_{\text{click}}$, is defined in Eq.~(\ref{eq:pclickBB84}). The
probability of a signal count, $p_{\text{signal}}$, is given by
Eq.~(\ref{eq:psignalBB84}), while the probability of a dark count,
$p_{\text{dark}}$, in this case is given by the expression:
\begin{equation}
p_{\text{dark}}=2d \label{eq:pdarkDPS}
\end{equation}
because there are two detectors at the receiver unit. Finally, the
error rate is defined in Eq.~(\ref{eq:eBB84}), and the values of
$f(e)$ are given in Table~\ref{tab:fe}.

In the case of small error rate and $p_{\text{dark}}\ll
p_{\text{signal}} \ll 1$, Eq.~(\ref{eq:RDPS}) gives
$R_{\text{DPSK}}\approx \nu (1-\frac{\mu}{N}) p_{\text{signal}}$
without a quantum memory, or $R_{\text{DPSK}}\approx \nu (1-2\mu)
p_{\text{signal}}$ with a quantum memory. This means that the secure
rate for the DPSK protocol decreases linearly with the fiber
transmission. This is in agreement with the results of \cite{gisin2}
and \cite{koashi}, who have considered a protocol similar to DPSK
and a slightly modified B92 protocol respectively.

\section{\label{sec:calc}Numerical results}
We compare the performance of quantum key distribution systems
implementing the BB84, BBM92 and DPSK protocols, when the
up-conversion single-photon detector is used. In order to do that,
we calculate the secure communication rate as a function of
distance for fiber-optic implementations of the three protocols,
based on Eqs.~(\ref{eq:RBB84}),(\ref{eq:RBBM92}) and
(\ref{eq:RDPS}) respectively. In the case of BB84 and BBM92, both
ideal and realistic sources of single and entangled photons are
considered. Some parameters are fixed in all simulations: the
channel loss is set to $\alpha = 0.2$ dB/km at 1.55 $\mu$m, the
baseline system error rate is set to $b = 0.01$, and in addition
to the fiber losses we assume an extra loss of $L_{\text{r}} = 1$
dB at the receiver site. As mentioned in Section~\ref{sec:BB84},
in the case of a weak laser pulse implementation of the BB84
protocol, the average number of photons per pulse, $\mu$, is an
adjustable parameter, with respect to which the rate is
numerically optimized at each distance. Intuitively, such
optimization is necessary because when this parameter is too low
the dark counts dominate, while when it is too high the
probability of multi-photon pulses becomes very large. In both
cases, secure communication quickly becomes impossible. The
corresponding adjustable parameter is $\chi$ in the case of the
BBM92 protocol with a Poissonian entangled-photon source.

%The rate is optimized with respect to $\mu$ in the case of the
%DPSK protocol as well,

It is clear from the analysis of Section~\ref{sec:rate} that the
critical parameters for the performance of a quantum cryptography
system related to the single-photon detector employed are the dark
counts per measurement time window, $d$, the quantum efficiency,
$\eta$, and the repetition rate of the transmission that it allows,
$\nu$. In the case of the up-conversion single-photon detector, due
to the non-gated mode operation of the Si APD there is no severe
limitation to the repetition rate of the experiment. In practice,
the limit is set by the speed of the electronic equipment as well as
by the timing jitter of the Si APD (typically $0.5-0.7$ nsec). A
realistic value, compatible with currently available components, is
$\nu_{\text{up}} = 1$ GHz. As was explained in Section~\ref{sec:up},
the limiting factor for the attainable communication rate is the
dead time of the Si APD, $t_{\text{d}}$. Assuming that the
photo-detection events follow a Poisson process, the probability of
two events occurring in a time period larger than $t_{\text{d}}$ is
given by the exponential factor $e^{-\delta\nu
p_{\text{click}}t_{\text{d}}}$, where $\delta$ depends on the number
of detectors in the receiver unit. For the typical value
$t_{\text{d}} = 50$ nsec, this saturation factor becomes rather
small at rates greater than a few MHz, limiting the final rate at
small fiber losses. Using Eqs.~(\ref{eq:etaup}) and (\ref{eq:dcup}),
we numerically optimize the communication rate for each protocol
with respect to the pump power, $p$, at each distance. Such
optimization is intuitively necessary because depending on the
communication distance an equilibrium between the values of the
quantum efficiency and the dark counts of the up-conversion detector
has to be established. The result of this optimization indicates the
optimal regime of operation of the detector at each distance.
Finally, the optimum filtering configuration, shown in
Fig.~\ref{fig:ideal-system}, is assumed, which sets the measurement
time window to 1 nsec.

\begin{figure}
\epsfxsize=3.25in \epsfbox{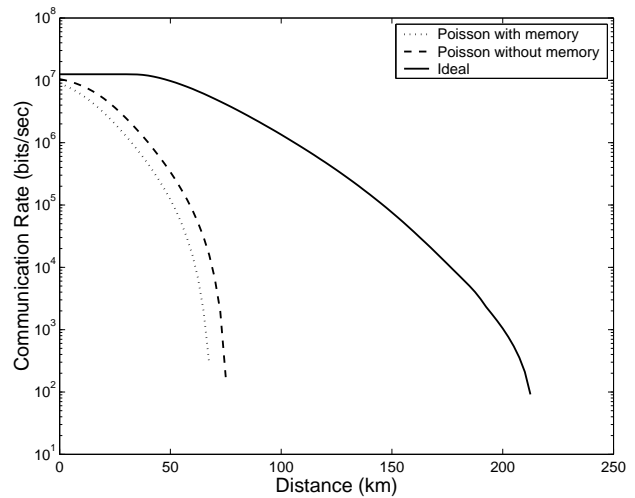} \caption{Secure communication
rate as a function of distance for the BB84 protocol employing a
Poisson or an ideal single-photon source.} \label{fig:BB84}
\end{figure}

\begin{figure}
\epsfxsize=3.25in \epsfbox{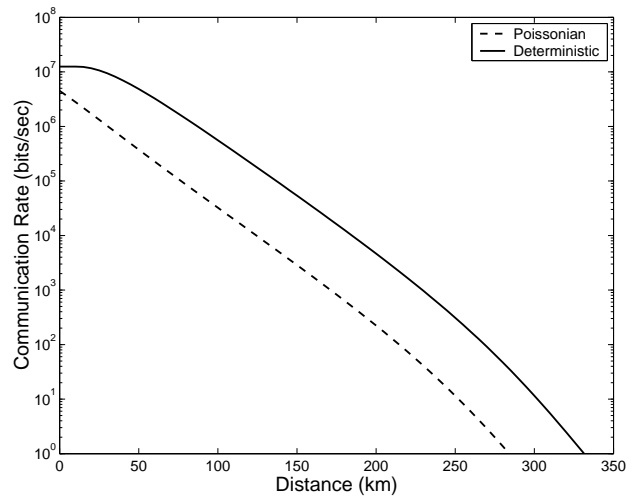} \caption{Secure communication
rate as a function of distance for the BBM92 protocol employing a
Poissonian or a deterministic entangled-photon source.}
\label{fig:BBM92}
\end{figure}

\begin{figure}
\epsfxsize=3.25in \epsfbox{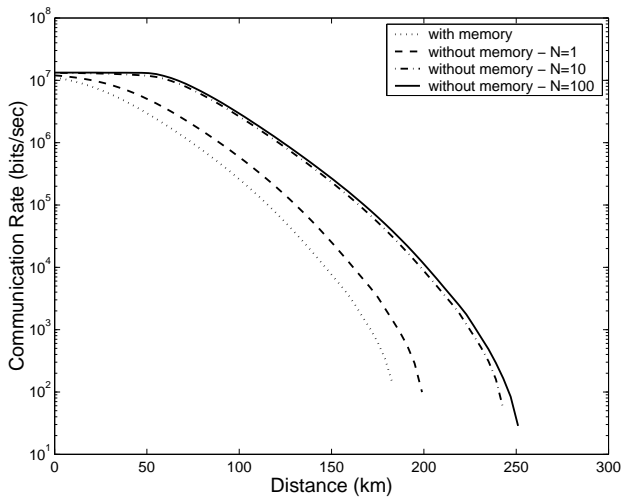} \caption{Secure communication
rate as a function of distance for the DPSK protocol employing time
delay parameters $N$ = 1, 10, or 100.} \label{fig:DPSK}
\end{figure}

The simulation results are shown in Figs.~\ref{fig:BB84},
\ref{fig:BBM92}, and \ref{fig:DPSK} for the BB84, BBM92, and DPSK
protocols respectively. Each curve features a cut-off distance,
which is due to the increasing contribution of the dark counts with
fiber length. The saturation effect, related to the dead time of the
Si APD, is apparent for small fiber losses and high bit rates.

In the case of BB84 with a Poisson single-photon source, we observe
in Fig.~\ref{fig:BB84} that not allowing Eve to possess a quantum
memory with an infinitely long coherence time does not have a major
effect on the performance of the system. The quadratic decrease of
the rate of the communication rate with the fiber length, a
consequence of the PNS attacks, is a dominant factor, making this
implementation unsuitable for long-distance quantum cryptography. On
the contrary, the use of an ideal single-photon source allows for a
significantly longer communication distance with high communication
rates. However, such a source does not exist today at 1.55 $\mu$m,
although efforts towards this goal are underway~\cite{fasel}.

As shown in Fig.~\ref{fig:BBM92}, the inherently more robust
entanglement-based BBM92 protocol allows for even longer
communication distances, having the capability to achieve a
practical 1 Hz secure key generation rate at more than 300 km with a
deterministic entangled-photon source. However, technological
difficulties related to entanglement generation and coincidence
detection at 1.55 $\mu$m have limited until today this distance to
30 km \cite{fasel2}.

The DPSK protocol features characteristics very similar to BB84 with
a single-photon source, due to its robustness to PNS attacks, as was
shown in the security analysis of Section~\ref{sec:DPS}. In this
case, when a realistic scenario is assumed, where Eve does not
possess a quantum memory with an infinitely long coherence time, or
Alice and Bob encrypt their public communication, we observe in
Fig.~\ref{fig:DPSK} a significant effect on the performance of the
system. Indeed, introducing a time delay parameter $N$ greater than
1 enhances both the secure communication rate and the communication
distance of the system considerably. Nevertheless, the advantage
becomes comparatively smaller as $N$ increases to values greater
than 10. This result shows that the DPSK protocol is a very
practical and appealing alternative for a long-distance QKD system,
with the potential of 1 kHz secure key generation rate over
distances longer than 200 km.

For all the QKD protocols, if instead of the up-conversion detector
we assume an InGaAs/InP APD with $\nu_{\text{APD}} = 10$ MHz, which
is the best gate frequency achieved to date \cite{yoshizawa}, and
the typical values $\eta_{\text{APD}} = 0.1$ and $d_{\text{APD}} =
10^{-5}$/gate \cite{gisin2}, we find that the maximum communication
distance is about half of the one achieved with an up-conversion
detector, while the communication rate is two orders of magnitude
lower than with the up-conversion detector, due to the gated-mode
operation of the InGaAs/InP APD. Clearly, the up-conversion detector
offers a great advantage over the InGaAs/InP APD as a single-photon
detector in a QKD system, both in terms of secure communication rate
and communication distance.

\begin{figure}
\epsfxsize=3.25in \epsfbox{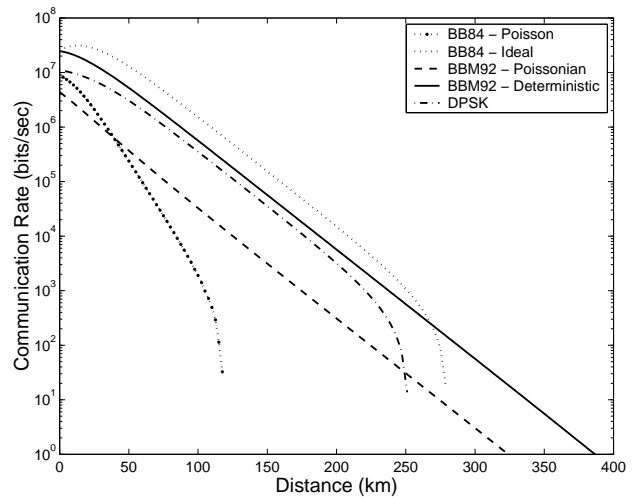} \caption{Comparison of the
performance of QKD systems implementing the BB84, BBM92 and DPSK
protocols. An optimized up-conversion single-photon detector and
Eve's ideal quantum memory are assumed in all cases.}
\label{fig:comparison}
\end{figure}

Finally, in Fig.~\ref{fig:comparison} we compare the performance of
quantum key distribution systems implementing the three protocols,
under the assumptions that Eve is equipped with an ideal quantum
memory and that the dark counts of the up-conversion detector,
caused by parasitic non-linear processes in the PPLN waveguide, are
eliminated. This means that the detector's performance is ideally
limited by the Si APD characteristics, which corresponds to
$d_{\text{up}}= 5\times 10^{-8}$. Operation at the maximum quantum
efficiency regime is also assumed, i.e. $\eta_{\text{up}}=0.46$. We
observe that, ultimately, 250 km of secure communication distance is
possible with the DPSK protocol and an ideal single-photon source
implementation of BB84, while BBM92 has the potential of extending
this distance to 350 km with a deterministic entangled-photon
source.

\section{\label{sec:concl}Conclusions}
In this paper, we studied the main characteristics of two types of
1.55 $\mu$m single-photon detectors, the InGaAs/InP APD and the
up-conversion detector, which combines frequency up-conversion in a
PPLN waveguide and detection by a silicon APD. We presented the
communication rate equations for the BB84 and the BBM92 QKD
protocols, and we derived a corresponding equation for the DPSK
protocol, developing a security analysis of this protocol against
certain types of hybrid attacks. Based on these equations, we
compared the performance of fiber-optic quantum key distribution
systems employing the protocols under consideration, with realistic
experimental parameters. In all cases, we found that a secure
communication rate of two orders of magnitude higher than before is
possible, while the use of the up-conversion detector enables
quantum key distribution over communication distances longer by a
factor of 2 than with an InGaAs/InP APD. Furthermore, the importance
of the implemented protocol was illustrated, and the impact of Eve's
allowed capabilities was investigated. We concluded that the simple
and efficient DPSK protocol allows for more than 200 km of secure
communication distance with high communication rates, in the
realistic case that Eve does not possess a quantum memory with an
infinitely long coherence time, and the time delay parameter $N$ is
greater than 1. The BBM92 protocol can extend this distance to 300
km with a reasonably high secure key generation rate. It is clear
that improving the performance of the Si APDs with respect to their
dead time and timing jitter and reducing the dark counts of the
up-converter will extend the capabilities of fiber-optic QKD systems
employing these protocols even further.

\begin{acknowledgments}
The authors would like to thank Edo Waks for his helpful comments
and suggestions. Financial support was provided by the MURI Center
for Photonic Quantum Information Systems (ARO/ARDA
DAAD19-03-1-0199), and the Quantum Entanglement Project, SORST,
JST.
\end{acknowledgments}

%\bibliography{ED-quantph-062005}

\end{document}